 \documentclass[prd,aps,
twocolumn,
nofootinbib,
floatfix,
superscriptaddress]{revtex4-2}
\usepackage{graphicx}
\usepackage{epsfig}
\usepackage{rotating}
\usepackage{amssymb}
\usepackage{subfigure}
\usepackage{dsfont}
\usepackage{psfrag}
\usepackage{bbold}
\usepackage{xcolor}
\usepackage{multirow}
\usepackage{amsmath,euscript,array,mathrsfs}

\newcommand{\beq}{\begin{equation}}
\newcommand{\eeq}{\end{equation}}
\newcommand{\beqs}{\begin{eqnarray}}
\newcommand{\eeqs}{\end{eqnarray}}

\newcommand{\Tr}{{\rm Tr}}

\newcommand{\SU}{\text{SU}}

\def\hbar{\hspace{0pt}\raisebox{1pt}{$-$} \hspace{-7pt} h}

\newcommand{\be}{\begin{equation}}
\newcommand{\ee}{\end{equation}}

\newcommand{\bea}{\begin{eqnarray}}
\newcommand{\eea}{\end{eqnarray}}

\def\lbldef#1#2{\expandafter\gdef\csname #1\endcsname {#2}}

\def\href#1#2{#2}

\newcommand{\ber}{\begin{eqnarray}}
\newcommand{\eer}{\end{eqnarray}}

\newcommand{\beqar}{\begin{eqnarray}}

\newcommand{\cO}{{\cal O}}

\newcommand{\eeqar}{\end{eqnarray}}

\newcommand{\dsl}
  {\kern.06em\hbox{\raise.15ex\hbox{$/$}\kern-.56em\hbox{$\partial$}}}

\newcommand{\eeqarr}{\end{eqnarray}}
\newcommand{\ZZ}{{\rm \kern 0.275em Z \kern -0.92em Z}\;}

\def\CC{{\mathchoice
{\rm C\mkern-8mu\vrule height1.45ex depth-.05ex
width.05em\mkern9mu\kern-.05em}
{\rm C\mkern-8mu\vrule height1.45ex depth-.05ex
width.05em\mkern9mu\kern-.05em}
{\rm C\mkern-8mu\vrule height1ex depth-.07ex
width.035em\mkern9mu\kern-.035em}
{\rm C\mkern-8mu\vrule height.65ex depth-.1ex
width.025em\mkern8mu\kern-.025em}}}

\def\RR{{\rm I\kern-1.6pt {\rm R}}}

\def\ZZ{{\rm Z}\kern-3.8pt {\rm Z} \kern2pt}
\def\IB{\relax{\rm I\kern-.18em B}}
\def\ID{\relax{\rm I\kern-.18em D}}
\def\II{\relax{\rm I\kern-.18em I}}
\def\IP{\relax{\rm I\kern-.18em P}}

\newcommand{\bear}{\begin{eqnarray}}
\newcommand{\eear}{\end{eqnarray}}

\def\cl{{closed}}

\def\6{\partial}

\def\bea{\begin{eqnarray}}
\def\eea{\end{eqnarray}}

\def\beqx{\begin{displaymath}}
\def\eeqx{\end{displaymath}}

\newcommand{\bmat}{\left(\begin{array}}
\newcommand{\emat}{\end{array}\right)}

\def\cl{{\cal L}}

\def\bo{{\raise-.3ex\hbox{\large$\Box$}}}               
\def\face{{\raise.2ex\hbox{$\displaystyle \bigodot$}\mskip-2.2mu \llap {$\ddot
        \smile$}}}                                   
\def\>{\rangle}                                      
\def\<{\langle}                                      

\def\VEV#1{\left\langle #1\right\rangle}             
\def\leftrightarrowfill{$\mathsurround=0pt \mathord\leftarrow \mkern-6mu
        \cleaders\hbox{$\mkern-2mu \mathord- \mkern-2mu$}\hfill
        \mkern-6mu \mathord\rightarrow$}        
\def\dvec#1{\vbox{\ialign{##\crcr
        \leftrightarrowfill\crcr\noalign{\kern-1pt\nointerlineskip}
        $\hfil\displaystyle{#1}\hfil$\crcr}}}           
\def\Tr{{\rm Tr \,}}                                    

\def\-{\hphantom{-}}

\begin{document}

\title{A Near-Conformal Composite Higgs Model}

\author{Thomas Appelquist}
\affiliation{Department of Physics, Sloane Laboratory, Yale University, New Haven, Connecticut 06520, USA}
\author{James Ingoldby}
\affiliation{Abdus Salam International Centre for Theoretical Physics, Strada Costiera 11, 34151, Trieste, Italy}
\author{Maurizio Piai}
\affiliation{Department of Physics, College of Science, Swansea University, Singleton Park, Swansea, Wales, UK}



\begin{abstract}
We analyze a composite Higgs model based on the confining $SU(3)$ gauge theory with $N_f = 8$ Dirac fermions in the fundamental representation. This gauge theory has been studied on the lattice and shown to be well described by a dilaton effective field theory (EFT). Here we modify the EFT by assigning standard-model quantum numbers such that four of the composite pseudo-Nambu-Goldstone boson (pNGB) fields form the standard-model Higgs doublet, by coupling it to the top quark, and by adding to the potential a term that triggers electroweak symmetry breaking. The model contains a pNGB Higgs boson, a set of heavier pNGBs, and an approximate dilaton in the same mass range. We study the phenomenology of the model, and discuss the amount of tuning required to ensure consistency with current direct and indirect bounds on new physics, highlighting the role of the dilaton field.
\end{abstract}

\maketitle


\section{Introduction}
\label{sec:Intro}
Lattice studies of the $SU(3)$ gauge theory with $N_f=8$ Dirac fermions in the fundamental representation show evidence of a light scalar singlet~\cite{Aoki:2014oha,Appelquist:2016viq,Aoki:2016wnc,Gasbarro:2017fmi,Appelquist:2018yqe}. (Similar results hold with $N_f=2$ fermions in the symmetric representation~\cite{Fodor:2012ty,Fodor:2015vwa,Fodor:2016pls,Fodor:2017nlp,Fodor:2019vmw,Fodor:2020niv}.) The suggestion that this state might be a dilaton has fueled a revival of interest in the dilaton effective field theory (EFT). Its history dates back to dynamical symmetry breaking~\cite{Leung:1985sn,Bardeen:1985sm,Yamawaki:1985zg}, well before this recent lattice-driven activity~\cite{Matsuzaki:2013eva,Golterman:2016lsd,Kasai:2016ifi,Hansen:2016fri, Golterman:2016cdd,Appelquist:2017wcg,Appelquist:2017vyy,Golterman:2018mfm,Cata:2019edh,Appelquist:2019lgk,Golterman:2020tdq,Golterman:2020utm}. Existing lattice data, analyzed via the dilaton EFT~\cite{Appelquist:2017wcg,Appelquist:2017vyy} yielded the first measurement of a key, large anomalous dimension related to the fermion bilinear condensate~\cite{Leung:1989hw}. The results are consistent with earlier expectations~\cite{Cohen:1988sq} and with recent high-loop perturbative studies~\cite{Ryttov:2010iz,Ryttov:2016asb}.

This theory, with a global $SU(8)\times SU(8)$ symmetry, broken to the diagonal $SU(8)$, is a natural candidate to build a composite Higgs 
model (CHM)~\cite{Kaplan:1983fs,Georgi:1984af,Dugan:1984hq}, (see also \cite{Vecchi:2015fma,Ma:2015gra,Panico:2015jxa,Panico:2015jxa,Witzel:2019jbe,Cacciapaglia:2020kgq,Ferretti:2016upr,Cacciapaglia:2019bqz} and references therein). Lattice studies of $SU(2)$~\cite{Hietanen:2014xca,Detmold:2014kba,Arthur:2016dir,Arthur:2016ozw,Pica:2016zst,Lee:2017uvl,Drach:2017btk}, 
$SU(4)$~\cite{Ayyar:2017qdf,Ayyar:2018zuk,Ayyar:2018ppa,Ayyar:2018glg,Cossu:2019hse}, and $Sp(4)$~\cite{Bennett:2017kga,Bennett:2019jzz,Bennett:2019cxd} 
gauge theories have explored the possible origin of CHMs. The $SU(3)$ gauge theory has  distinctive features: the presence of a light scalar singlet which modifies the EFT description of the CHM (see also Ref.~\cite{BuarqueFranzosi:2018eaj}), and the presence of large anomalous dimensions. Furthermore, ordinary baryons can give rise to top compositeness~\cite{Vecchi:2015fma}.

In this paper, we show that the presence of the dilaton field in the EFT allows us to construct an appealing CHM based upon the $SU(8)\times SU(8)/SU(8)$ coset. 
We demonstrate that observables such as the ratio of the mass of the Higgs boson, $m_h\simeq 126$ GeV, to the electroweak vacuum expectation value (VEV), $v\simeq 246$ GeV, and to the mass of the additional heavy scalars, are substantially altered with respect to generic CHM expectations. We highlight how current lattice studies might already be exploring phenomenologically relevant regions of parameter space. These statements depend on the value of a (currently) unknown scaling dimension $w$, which in principle can be measured on the lattice.

\section{The Model}
\label{sec:Model}

\begin{table}[t]
	\vspace{10pt}
	\centering
	\renewcommand\arraystretch{1.2}
	\begin{tabular}{| c |  c  c  c | c |}
		\hline\hline
		Fermion
		 & $\SU(2)_L$ & $\text{U}(1)_Y$ & $SU(3)_c$& $SU(3)$\\
		\hline
		$L_\alpha$  & 2 & 0 & 1 & 3\\
		$R_{1,2}$  & 1 & $\begin{pmatrix} 1/2\\ -1/2 \end{pmatrix}$ & 1 & 3 \\
		$T$  & 1 & 2/3 & 3 & 3  \\
		$S$  & 1 & 0  & 1 & 3 \\
		\hline\hline
	\end{tabular}
	\caption{Quantum number assignments of the Dirac fermions. $SU(3)_c \times SU(2)_L \times U(1)_Y$ 
		is the SM gauge group, while $SU(3)$ is the strongly coupled gauge symmetry. We denote with $\alpha=1,\,2$ the $SU(2)_L$ index.
		The fermions denoted by  $R_{1,2}$ form
		a fundamental representation of the global $SU(2)_R$ custodial symmetry. A model with similar
		assignments has been considered in Ref.~\cite{Vecchi:2015fma}.}\label{Tab:gauge}
\end{table}

We assign to the eight Dirac fermions the quantum numbers indicated in Table~\ref{Tab:gauge}.
The global $SU(8)\times SU(8)$ symmetry group is broken both explicitly (by a diagonal mass term) and spontaneously
(by the strong dynamics) to its diagonal $SU(8)$ subgroup. The gauge group
of the standard model (SM) is a subgroup of the unbroken $SU(8)$.
The EFT description contains $63$ pseudo-Nambu-Goldstone Bosons (pNGBs) denoted as $\pi^a$, and one additional $SU(8)$ singlet, the dilaton, which we denote as $\chi$. We ignore the $U(1)_A$ meson, which is a singlet and has a large mass, due to the anomaly.

The CHM construction starts from the observation that $8$ of the pNGBs have the correct quantum numbers to form two copies of the Higgs doublet of the standard model. We further modify the dilaton EFT of Refs.~\cite{Appelquist:2017wcg,Appelquist:2017vyy,Appelquist:2019lgk} by adding two terms: a coupling of one of these two doublets to the top quark, and a related potential term for the pNGBs. In this paper, we ignore all SM fermions other than the top; the generalization to include other SM fermions within the EFT framework is straightforward.

The EFT Lagrangian density that results from this construction is the following:
\begin{align}
	\cl =  \frac{1}{2}\left(\partial_\mu\chi\right)^2 & +\cl_\pi + \cl_M - V(\chi)\nonumber\\  +& \cl_Y - V_t+\cl_1,\label{eq:LOEFT}
\end{align}
where the dilaton field $\chi$ acts as a conformal compensator, coupling to EFT operators in such a way as to restore scale invariance in Eq.~(\ref{eq:LOEFT}). It acquires a VEV $\VEV{\chi}\equiv F_d$, breaking scale invariance spontaneously.

The kinetic term for the pNGBs is
\begin{align}
\cl_\pi = \frac{F^2_\pi}{4}\left(\frac{\chi}{F_d}\right)^2\Tr\left[D_\mu\Sigma\left(D^\mu\Sigma\right)^\dagger\right],
\label{eq:Lkin}
\end{align}
where $F_\pi$ enters the EFT as the scale of spontaneous breaking of $SU(8)\times SU(8)$. The matrix--valued field $\Sigma$ represents the 63 pNGBs spanning the $SU(8)\times SU(8)/SU(8)$ coset. The covariant derivatives describe couplings to the SM gauge bosons, following the embedding identified in Table~\ref{Tab:gauge}. Their kinetic terms and self interactions are the standard ones, which we include in $\cl_1$. $\Sigma$ satisfies the nonlinear constraint $\Sigma\Sigma^\dagger=\mathbb{1}_8$.

The Dirac mass given to the fermions of the new strong sector leads directly to the following term in the EFT 
\begin{align}
\cl_M = \frac{M^2_\pi F^2_\pi}{4}\left(\frac{\chi}{F_d}\right)^y\Tr\left[\Sigma+\Sigma^\dagger\right],\label{eq:Lm}
\end{align}
and breaks the global symmetry. The quantity $M^2_\pi$ sets the scale for the masses of the 59 pNGBs besides those that become the Higgs doublet. The parameter $y$ has been interpreted as the scaling dimension of the fermion bilinear condensate in Ref.~\cite{Leung:1989hw}. Its value is $y=2.06\pm0.05$ \cite{Appelquist:2019lgk}.

The scalar potential $V(\chi)$ describes the self-interactions of the dilaton field. It encodes both the spontaneous and explicit breaking of scale symmetry originating from the underlying gauge theory. We provided a general form for this potential in Ref.~\cite{Appelquist:2019lgk}, where it played a key role. Here we will not find it necessary to further invoke the explicit form of $V(\chi)$.

At the level of the EFT, we describe the mass of the top quark using
the Yukawa interaction
\begin{align}
\cl_Y = y_tF_\pi\left(\frac{\chi}{F_d}\right)^{z}\left(\bar{Q}^\alpha_L t_R\right)\Tr\left[P_\alpha\Sigma\right] + \text{h.c.}\,.\label{eq:ETCYukawa}
\end{align}
The underlying gauge theory determines the scaling dimension $z$. $H_{\alpha}\equiv Tr[P_{\alpha} \Sigma]$ transforms as the Higgs Doublet, with quantum numbers $(2,-1/2)$ under $SU(2)_L\times U(1)_Y$. Here $\alpha$ is the index of $\SU(2)_L$. We take the projectors $P_\alpha$ to be the following $8\times8$ matrices
\begin{align}
P_\alpha=\begin{pmatrix}
\tilde{P}_\alpha & \mathbb{0}_{4}\\
\mathbb{0}_{4} & \mathbb{0}_{4}
\end{pmatrix},
\label{eq:proj}
\end{align}
with
\begin{align} 
\tilde{P}_1=\frac{1}{2}\begin{pmatrix}
0 & 0 & 1 & 0  \\
0 & 0 & 0 & 0  \\
0 & 0 & 0 & 0  \\
0 & -1 & 0 & 0  \\
\end{pmatrix},\quad
\tilde{P}_2=\frac{1}{2}\begin{pmatrix}
0 & 0 & 0 & 0  \\
0 & 0 & 1 & 0  \\
0 & 0 & 0 & 0  \\
1 & 0 & 0 & 0  \\
\end{pmatrix}.\label{eq:Pmatrix}
\end{align}
This choice, motivated by simplicity, will ensure that one Higgs doublet is responsible for electroweak symmetry breaking.

The operator in Eq.~(\ref{eq:ETCYukawa}) breaks the $SU(8)\times SU(8)$ global symmetry. In the underlying theory, interactions responsible for generating Eq.~(\ref{eq:ETCYukawa}) also generate an $SU(8)\times SU(8)$ breaking contribution to the potential of the form
\begin{align}
V_t=- C_t\left(\frac{\chi}{F_d}\right)^{w}\;\sum_{\alpha=1}^{2}\left|\Tr\left[P_\alpha\Sigma\right]\right|^2.
\label{eq:Vt}
\end{align}
The (unknown) scaling dimension $w$ derives from the underlying gauge theory. This potential has non-Abelian global symmetry $SU(2)_L\times SU(2)_R\times SU(4)$, the $SU(4)$ remaining due to the vanishing entries in $P_\alpha$. The custodial $SU(2)_R$ symmetry, suppressing the effect of new physics on precision electroweak observables, is preserved here for our choice of $P_\alpha$, despite the fact that the Yukawa interaction in Eq.~(\ref{eq:ETCYukawa}) breaks this symmetry explicitly. There are subleading contributions to this potential that break $SU(2)_R$, but they are smaller than $V_t$, and we will not consider them. Similarly, the gauging of the SM subgroup breaks the global symmetries, and leads to additional contributions to the potential, but they are smaller and we neglect them in this analysis. 

A loop of top quarks can also generate the interaction in Eq.~(\ref{eq:Vt}), naturally making the constant $C_t$ positive. In addition, partial top compositeness can generate Eq.~(\ref{eq:ETCYukawa})---and hence Eq.~(\ref{eq:Vt})---in which case the field $Q^\alpha_L$ couples linearly to the baryon operator ${\cal B}^\alpha_L=L^\alpha(\overline{TR_2})^\dagger+R_2(\overline{TL})^{\dagger\,\alpha}$ and $t_R$ to ${\cal B}_R=2R_1(TR_2)+L(TL)$ in the underlying theory. (Parentheses indicate contraction of spinor indices.) Alternatively, Eq.~(\ref{eq:ETCYukawa}) can be generated by coupling the elementary fermion bilinear $\left(\bar{Q}^\alpha_L t_R\right)$ to the mesonic operator $\cO_M^\alpha=(L^\alpha\overline{R}_1)-\epsilon^{\alpha\beta}\left(R_2\overline{L}_\beta\right)$.

\section{The Vacuum}
\label{sec:Vacuum}

We first analyze the vacuum of the EFT. Both the pNGB that we identify with the Higgs field and the dilaton have nontrivial vacuum values, which  must be calculated simultaneously. Then we determine the mass of the composite Higgs boson in this vacuum, emphasizing the significant role played by the dilaton field.

The three terms without derivatives or the top quark field in the Lagrangian of Eq.~(\ref{eq:LOEFT}) define a potential for both the pNGBs and the dilaton. It is helpful to parametrize the $\Sigma$ field as 
\begin{align}
\Sigma=\exp{\left[i\theta\begin{pmatrix}
	\mathbb{0}_{2\times 2} & -i\mathbb{1}_2 & \mathbb{0}_{2\times 4}\\
	i\mathbb{1}_2 & \mathbb{0}_{2\times 2} & \mathbb{0}_{2\times 4} \\
	\mathbb{0}_{4\times 2} & \mathbb{0}_{4\times 2} & \mathbb{0}_{4\times 4}
	\end{pmatrix}\right]},\label{eq:vacdirection}
\end{align}
where only the degree of freedom corresponding to the pNGB component of the Higgs boson (represented by $\theta$) is shown, for simplicity. The potential then reads
\begin{multline}
W(\chi,\theta)=V(\chi)-C_t\left(\frac{\chi}{F_d}\right)^w\sin^2\theta\\-2M^2_\pi F^2_\pi\left(\frac{\chi}{F_d}\right)^y\left(1+\cos\theta\right),
\label{eq:Vctheta}
\end{multline}
Minimizing this potential determines the vacuum value $F_d$ of $\chi$, and the vacuum value of $\theta$ (the misalignment angle). We henceforth use $\theta$ to denote this vacuum value rather than the dynamical pNGB field. The electroweak scale $v\simeq246$ GeV is related to the misalignment angle through $v=\sqrt{2}F_\pi\sin\theta$. The top acquires the mass $m_t=y_tv/\sqrt{2}$.

At the minimum of the potential, we have $\langle\chi\rangle\equiv F_d$, while
\begin{align}
\cos\theta = \frac{M^2_\pi F^2_\pi}{C_t},
\label{eq:misalignment}
\end{align}
provided that $M_\pi^2F_\pi^2 < C_t$, otherwise at the minimum we get $\theta=0$, preserving electroweak symmetry. Furthermore, the minimum must satisfy:
\begin{multline}
0=\left.\frac{\partial V}{\partial \chi}\right|_{F_d}-\frac{4yM^2_\pi F^2_\pi}{F_d}\\-\frac{M^2_\pi F^2_\pi}{F_d}\left(w\frac{\sin^2\theta}{\cos\theta}-2y(1-\cos\theta)\right).\label{eq:Fdmin}
\end{multline}
These equations determine $\theta$ in terms of $C_t$ and provide a relation between $F_d$ and other EFT parameters with the dilaton potential $V$. 

To comport with the SM at currently accessible energies, we must find a small misalignment angle $\theta\ll1$, that is, a large separation between $v$ and $F_{\pi}$. This is achieved by tuning $C_t$ in Eq.~(\ref{eq:misalignment}).

For $\theta \ll 1$, Eq.~(\ref{eq:Fdmin}) determining $F_d$ simplifies in an essential way. The second line is suppressed, and may be neglected in first approximation. The resulting equation is precisely the one used in Refs.~\cite{Appelquist:2017wcg,Appelquist:2017vyy} to relate $F_d$ to the other parameters in the EFT employed there. That EFT, with no potential term proportional to $C_t$, was used to fit lattice data for the $SU(3)$ gauge theory with $N_f = 8$. The functional form of the scalar potential $V(\chi)$ was constrained in that fit.

The mass matrix for the $\chi$ and $\theta$ degrees of freedom is approximately given (for small misalignment angle) by
\begin{align}
{\cal M}^2=
\begin{pmatrix}
M^2_d & \theta\frac{\sqrt{2}M^2_\pi F_\pi(y-w)}{F_d}\\\theta\frac{\sqrt{2}M^2_\pi F_\pi(y-w)}{F_d} & \theta^2 M^2_\pi 
\end{pmatrix},
\label{eq:massmatrix}
\end{align}
The (1,1) entry is the second derivative of $W(\chi,\theta)$ with respect to $\chi$ at $\chi=F_d$ in the limit $\theta \rightarrow 0$. It is expressible in terms of the scalar
potential $V(\chi)$ and other EFT parameters by
\begin{align}
M^2_d = \left.\frac{\partial^2V}{\partial\chi^2}\right|_{F_d}-4y(y-1)\frac{F^2_\pi}{F^2_d}M^2_\pi,
\end{align}
which was employed in Refs.~\cite{Appelquist:2017wcg,Appelquist:2017vyy} to fit lattice data. In the present context, $M_d$ is the approximate mass of the heavy scalar eigenstate, composed principally of $\chi$.

By diagonalizing the mass matrix in Eq.~(\ref{eq:massmatrix}), we find that the mass $m_h$ of the lightest eigenstate (corresponding to the Higgs boson) is given by
\begin{align}
\frac{m^2_h}{v^2}= \frac{M^2_\pi}{2F^2_\pi}\left(1-\frac{2M^2_\pi F^2_\pi(y-w)^2}{M^2_d F^2_d}\right),\label{eq:mh}
\end{align}
up to $O(\theta)$ corrections. The second term in the parentheses, arising from the presence of the dilaton field, is a distinctive feature of this model. Its presence allows us to accommodate the measured ratio $m_h^2/v^2 \approx 0.26$, drawing directly on lattice data for which $M_{\pi}^2/2F_{\pi}^2$ is typically an order of magnitude larger.

\section{Phenomenology}
\label{sec:Pheno}

In this section we examine the spectrum of particles of our CHM and the conditions under which constraints from collider experiments are satisfied. We also discuss the amount of fine tuning needed for the model to satisfy the experimental constraints. 

The spin-0 part of the spectrum of our EFT consists of $63$ NGBs and pNGBs associated with the spontaneous breaking of the $SU(8)\times SU(8)$ symmetry of the underlying gauge theory, along with a scalar state of approximate mass $M_d$. The 3 massless NGBs are eaten by the $W^{\pm}$ and $Z$. One state is the relatively light pNGB Higgs boson of mass $m_h$ (Eq.~(\ref{eq:mh})), while $59$ are heavier pNGB states with their mass scale set by $M_{\pi}$. One additional heavier state has mass $M_d$. The quantities $F_{\pi}$ and $F_d$ are decay constants associated with these states. To set the relative size of $M_{\pi}$, $M_d$, $F_{\pi}$, we draw directly from the LSD lattice measurements~\cite{Appelquist:2018yqe}.

Neglecting the SM gauge interactions, the EFT has approximate $SU(2)_L\times SU(2)_R\times SU(4)$ global symmetry. The pNGB multiplet decomposes into representations of this symmetry as follows
\beqs
63&=& (3,1,1)+(1,3,1)+(1,1,1)+(2,2,1)\nonumber\\
&&+(1,1,15)+(2,1,4)+(1,2,4)+(1,1,1)
\,.
\eeqs
The misaligned vacuum breaks $SU(2)_L\times SU(2)_R$ spontaneously to its diagonal subgroup $SU(2)_D$. As a result, the composite spectrum is organized in a set of (approximate) multiplets of $SU(2)_D\times SU(4)$.

To determine the spectrum, we first specify the quantities  $\{M^2_\pi,M^2_d,F^2_\pi,F^2_d,y,m_h,v\}$. We use data from lattice studies of the $N_f=8$ gauge theory. We extract the ratios $M^2_\pi/F^2_\pi$ and $M^2_d/F^2_\pi$ from Tables. III and IV of Ref.~\cite{Appelquist:2018yqe} for five different constituent fermion masses $m_{fi}$. We then take $y=2.06\pm0.05$ and $F^2_\pi/F^2_d=0.086\pm0.015$ from Ref.~\cite{Appelquist:2019lgk}, and set $m_h$ and $v$ to their experimentally determined values. Finally, we must set the overall scale for the new composite sector. As a benchmark, we take $M_\pi=4\,\,\text{TeV}$, to ensure that the 59 heavier pNGBs lie outside the reach of direct searches. The strongest bounds coming from searches for color octet scalars indicate that $M_{\pi}\gtrsim3.7$ TeV~\cite{Sirunyan:2019vgj}, although the precise bound depends on an additional coupling~\cite{Belyaev:2016ftv,Cacciapaglia:2020vyf}. The calculated spectrum is then shown in Table~\ref{Tab:spectrum}. It is nearly independent of which lattice point ($m_{fi}$ value) we use in the analysis. The small variation in the mass of the heaviest singlet state (the dilaton) is mostly due to fluctuations in the lattice measurement of $M^2_d/M^2_\pi$.

In Table~\ref{Tab:spectrum} we show only the central values for the masses of the spin--0 states, for the representative choices of parameters discussed previously, in order to illustrate the high degree of degeneracy among the heavier states, all of which have masses sitting within 15\% of one another. The overall determination of the mass scale is affected by the large (and correlated) uncertainties originating in the lattice measurements of $M_d$ and $F_\pi$, and the indirect determination of $F_d$.

	\begin{table}[t]
		\vspace{10pt}
		\centering
		\setlength{\tabcolsep}{5pt}
		\renewcommand\arraystretch{1.2}
		\begin{tabular}{| c  c | c  c  c  c  c |}
			\hline\hline
		 	\multirow{2}{*}{$SU(2)_D$} & \multirow{2}{*}{$SU(4)$} & \multicolumn{5}{c|}{Mass (TeV)}\\
		     & & $m_{f1}$ & $m_{f2}$ & $m_{f3}$ & $m_{f4}$ & $m_{f5}$ \\
			\hline
			 1 & 1 & 4.31 & 4.73 & 4.29 & 4.96 & 4.87 \\
			 3 & 1 & 4.35 & 4.37 & 4.39 & 4.40 & 4.40 \\
			 2 & 4 & 4.18 & 4.19 & 4.20 & 4.20 & 4.20 \\
			 3 & 1 & 4.03 & 4.03 & 4.04 & 4.04 & 4.04 \\
			 1 & 1 & 4.03 & 4.03 & 4.04 & 4.04 & 4.04 \\
			 1 & 1 & 4.03 & 4.03 & 4.04 & 4.04 & 4.04 \\
			 1 & 15 & 4.00 & 4.00 & 4.00 & 4.00 & 4.00 \\
			 1 & 1 & 3.99 & 3.98  & 3.98 & 3.98 & 3.98 \\
			 2 & 4 & 3.84 & 3.83 & 3.83 & 3.82 & 3.82 \\
			 3 & 1 & 3.67 & 3.66 & 3.64 & 3.64 & 3.64 \\
			 1 & 1 & 0.126 & 0.126 & 0.126 & 0.126 & 0.126 \\
			 3 & 1 & 0 & 0 & 0 & 0 & 0\\
			\hline\hline
		\end{tabular}
		\caption{Estimated composite spectrum determined using the procedure described in the text. The states are labeled using their $SU(2)_D\times SU(4)$ quantum numbers, shown in the left-hand column. The $m_{fi}$ refer to the 5 different values for the constituent fermion mass appearing in the lattice study of Ref.~\cite{Appelquist:2018yqe} (arranged in ascending order) that is used as an input into these estimates. We show only the central values of the masses, to highlight the high level of degeneracy among the heavier states, which is independent of uncertainties coming from the lattice determinations of $M^2_d$ and $F^2_\pi/F^2_d$.} \label{Tab:spectrum}
	\end{table}

In calculating the spectrum, the quantities $w$ (for which there is no existing determination from lattice data) and $C_t$ are chosen to reproduce the aforementioned constraints. We find that $C_t\approx(2\;\text{TeV})^4$, with the precise determination depending on the lattice point considered. Similarly, $F_\pi\approx1\;\text{TeV}$.

\begin{figure}
	\centering
	\begin{picture}(210,150)
	\put(10,5){\includegraphics[width=.38\textwidth ]{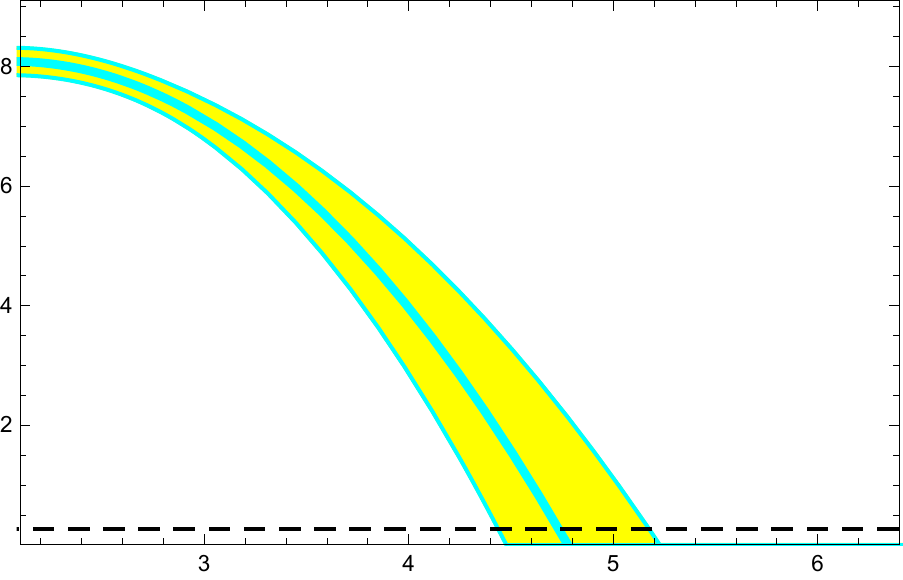}}
	\put(0,134){$m_h^2/v^2$}
	\put(190,0){$w$}
	\end{picture}
	\vspace{0mm}
	\caption{The ratio $m_{h}^2/v^2$, as a function of $w$, for $y=2.06\pm 0.05$ 
		and $F_{\pi}^2/F_{d}^2=0.086\pm 0.015$~\cite{Appelquist:2019lgk}, for the LSD 
		measurements taken at the second fermion mass ($m_{f2}$) in the range~\cite{Appelquist:2018yqe}. 
		The horizontal black dashed line represents the experimental value 
		$m_{h}^2/v^2\simeq 0.26$. The yellow shaded region is the uncertainty, which is dominated by the substantial uncertainty
		in the measurement of the mass $M_d$ of the scalar,
		and in the quantity $F_{\pi}^2/F_{d}^2$. For the best case scenario of $w=4.7$, the resulting uncertainty in $m^2_h/v^2$ is approximately an order of magnitude larger than the central value. Still, as the error in the measurement of $M_d$ is reduced, the range of acceptable $w$ values required to maintain $m_h^2/v^2 \approx 0.26$ will diminish and potentially shift slightly.}
	\label{Fig:lattice}
\end{figure}

In Fig.~\ref{Fig:lattice}, we show the impact of uncertainties in the lattice data on the allowed range for the scaling dimension $w$. For illustration purposes, we take lattice data for $M^2_\pi/M^2_d$ at the second fermion mass point $m_{f2}$ from Ref.~\cite{Appelquist:2018yqe} (along with values for $y$ and $F^2_\pi/F^2_d$ from Ref.~\cite{Appelquist:2019lgk}) accounting for their uncertainties and shade in yellow the allowed ranges for $m^2_h/v^2$ and $w$. We see that if we require $m^2_h/v^2\simeq0.26$ for consistency with experiment, $w$ could lie anywhere in the range $4.5<w<5.2$. Similar results hold for the other Table~\ref{Tab:spectrum} values of the fermion mass $m_f$. If lattice simulations are able to measure $w$ with some precision (and measure $M_{\pi}^{2}F_{\pi}^{2}/M_d^{2}F_d^{2}$ with a similar precision), then meeting the requirement that $m_h^2/v^2\simeq 0.26$ could require an $m_f$ value outside the range of Table~\ref{Tab:spectrum}.

Obtaining a spectrum with a realistic hierarchy $m_h\ll M_\pi$ does require tuning $C_t$. For the parameter choices required to produce the spectrum in Table~\ref{Tab:spectrum}, the misalignment angle satisfies $\theta\simeq0.19$, implying a tuning for $C_t$ of order 2\% through Eq.~(\ref{eq:misalignment}). The dilaton indirectly reduces the requisite tuning through Eq.~(\ref{eq:mh}): the effect of the term in parentheses (which depends on the dilaton) is to further suppress $m^2_h/M^2_\pi$, helping the colored pNGBs evade direct detection bounds.

Couplings between the Higgs and pairs of W or Z gauge bosons, as well as couplings between the Higgs and pairs of top quarks, deviate from their SM predictions in this CHM. These couplings include contributions from both the pNGB and dilaton components of the Higgs particle, and the expressions for them take the same form as those derived in the literature (see e.g. Ref.~\cite{Cacciapaglia:2020kgq} and refs. therein). In the limit $\theta\rightarrow 0$, the Higgs would couple to the gauge bosons and top with the same strength as the SM Higgs.

Using the benchmark $M_\pi=4$ TeV and the values for parameters selected by the lattice data of Ref.~\cite{Appelquist:2018yqe}, we find that the ratio between the Higgs couplings to W and Z bosons and their SM values is approximately $0.98$. The coupling to top pairs has additional weak dependence on the unknown scaling dimension $z$, which arises because the dilaton component of the Higgs boson has a coupling to the top that is $z$ dependent from Eq.~(\ref{eq:ETCYukawa}). For the value $z\simeq2-3$, the coupling strength to the top becomes the same as in the SM. Over a plausible range of values for $z$, the top coupling deviates from its SM value only by a few percent.

The amplitudes $h\rightarrow gg$ and $h\rightarrow\gamma\gamma$ also deviate slightly from their SM values. New electrically charged and colored pNGBs contribute to these amplitudes at loop level, but are sufficiently heavy for our choice of benchmark that their contributions are negligible. 

Considering all of these deviations, the signal significance for Higgs boson production in all observable channels would deviate from the SM prediction by no more than a few percent. Given the current accuracy of the Higgs measurements, which is no better than 8\%~\cite{Aad:2019mbh}, these effects will lie within experimental bounds, and a more precise analysis can be deferred.

The masses of composite states which are not included in the EFT have also been calculated in the $N_f=8$ gauge theory on the lattice, in Ref.~\cite{Appelquist:2018yqe}. In particular, this data allows us to estimate the masses of the vector ($\rho$) and axial ($a_1$) mesons. Using lattice measurements for ratios $M_\rho/M_\pi$ and $M_{a_1}/M_\pi$, we estimate that the $\rho$ would have a mass in the 6--8 TeV range and the $a_1$ a mass in the 9--11 TeV range, for our choice of benchmark $M_\pi=4\,\,\text{TeV}$. We therefore do not expect them to be detectable at the LHC. Given the small deviations in Higgs couplings, as well as the large $\rho$ and $a_1$ masses, precision electroweak observables such as the S parameter will lie within current experimental bounds.

\section{Summary}
\label{sec:Summary}

We have argued that the $SU(3)$ gauge theory with $N_f=8$ fundamental fermions provides an attractive ultraviolet completion for a realistic composite Higgs model. This model has the distinctive feature that the near--conformal behavior of its underlying dynamics has been revealed by lattice studies.

We have drawn on such lattice results to compute several observable quantities within the model. These include the misalignment angle in the vacuum of the theory, the mass of the Higgs boson, and the spectrum of heavy scalars. We have also examined the Higgs boson couplings and its production rates at the LHC. The model passes all the direct and indirect tests currently available, at the price of a moderate amount of fine-tuning for one of the coefficients in the EFT potential.

We have described the model in terms of the dilaton effective field theory (EFT) from Refs.~\cite{Appelquist:2017vyy,Appelquist:2017wcg,Appelquist:2019lgk}, requiring only a simple addition to realize the Higgs doublet as composite pseudo Nambu-Goldstone bosons. Because of the approximate scale invariance of the dilaton EFT, it is possible to accommodate a realistic value of the mass of the composite Higgs boson even for the values of the ratio $M_{\pi}/F_{\pi}$ currently available from lattice studies. As a consequence, the mass of the Higgs boson is suppressed by an order of magnitude with respect to that of the other pNGBs and dilaton in the EFT.

We look to future lattice studies for a determination of the scaling dimensions $z$ and $w$, which play important phenomenological roles. It will also be interesting to perform a more detailed study of the precision electroweak observables and explore the rest of the parameter space of this theory.

\vspace{0.5cm}
\begin{acknowledgments}
	
	MP would like to thank G.~Ferretti for a useful discussion.
	The work of MP has been supported in part by the STFC Consolidated Grants ST/P00055X/1 and ST/T000813/1. MP has also received funding from the European Research Council (ERC) under the European Union's Horizon 2020 research and innovation programme under grant agreement No 813942.

\vspace{1.0cm}

\end{acknowledgments}

\end{document}